\documentclass[letterpaper, 11pt]{article}
\usepackage{graphicx}
\usepackage{natbib}
\usepackage[left=2.5cm,top=2.5cm,right=2.5cm]{geometry}
\usepackage{parskip}
\usepackage{amsmath}
\usepackage{dsfont}

\newcommand{\hyperparams}{\boldsymbol{\alpha}}
\newcommand{\xx}{\mathbf{x}}

\title{Inference for Trans-dimensional Bayesian Models with Diffusive Nested
Sampling}
\author{Brendon J. Brewer}
\date{Department of Statistics, The University of Auckland\\
Private Bag 92019, Auckland 1142, New Zealand\\
\vspace{0.3cm}
{\tt bj.brewer@auckland.ac.nz}}

\begin{document}
\maketitle
\abstract{Many inference problems involve inferring the number $N$ of
components in some region, along with their properties $\{\xx_i\}_{i=1}^N$,
from a dataset $\mathcal{D}$. A common statistical example is finite mixture
modelling. In the Bayesian framework, these problems are typically solved using
one of the following two methods: i)
by executing a Monte Carlo algorithm (such as Nested Sampling) once for each
possible value of $N$,
and calculating the marginal likelihood or evidence as a function of $N$; or
ii) by doing a single run that allows the model dimension $N$ to change (such as
Markov Chain Monte Carlo with birth/death moves), and obtaining the posterior
for $N$ directly. In this paper we present a general approach to this problem
that uses trans-dimensional MCMC embedded {\it within} a Nested Sampling
algorithm, allowing us to explore the posterior distribution and calculate
the marginal likelihood (summed over $N$) even if the problem contains a phase
transition or other difficult features such as multimodality.
We present two example problems, finding sinusoidal signals in
noisy data, and finding and measuring galaxies in a noisy astronomical image.
Both of the examples demonstrate phase transitions in the relationship between
the likelihood and the cumulative prior mass, highlighting the need for
Nested Sampling.}

\section{Introduction}\label{sec:introduction}
\setlength{\parindent}{0cm}
\setlength{\parskip}{3mm}
Consider the following class of inference problems. There is some unknown
number, $N$, of components in a (physical or metaphorical) region. Each of the $N$ components has parameters
$\xx$, which may be a single scalar value (for example, a mass), or a set of
values (e.g. a two-dimensional position and a mass). These problems are often
challenging to solve due to the unknown (and potentially large) dimensionality.
In addition, they also have the so-called ``label-switching degeneracy'' issue
\citep{label_switching}, where the meaning of the model is invariant to
switching the identities of the components. Finite mixture modelling is a common
statistical example, but this kind of problem appears in many other contexts
\citep[e.g.][]{massinf}.

To carry out the inference, we must assign a prior to the components' parameters
$\{\xx_i\}_{i=1}^N$.
Since $N$ may be large, it is usually easier to assign an ``conditional prior''
given some hyperparameters $\hyperparams$, and then assign a prior to
$\hyperparams$. This kind of model is usually called {\it hierarchical}.
The prior for $N$, $\hyperparams$, and $\{\xx_i\}$ is usually factorised
in the following way:

\begin{eqnarray}
p(N, \hyperparams, \{\xx_i\}) &=& p(N) p(\hyperparams | N) p(\{\xx_i\} | \hyperparams, N) \\
&=& p(N) p(\hyperparams) \prod_{i=1}^N p(\xx_i | \hyperparams).
\end{eqnarray}

\begin{figure}
\begin{center}
\includegraphics{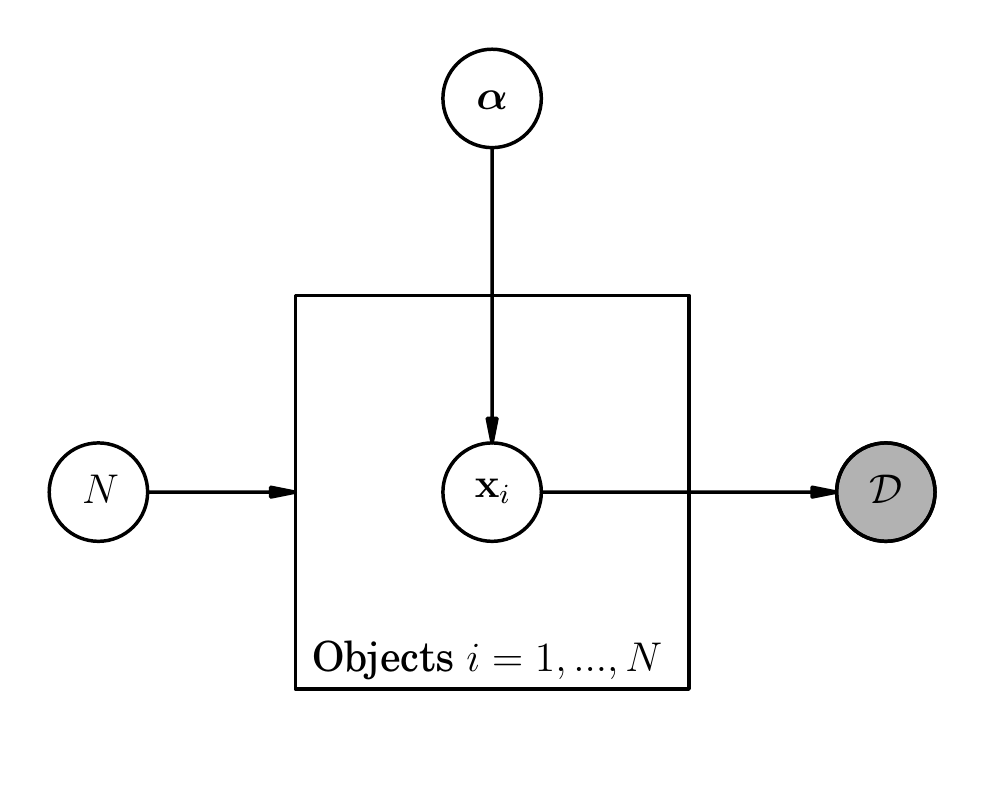}
\caption{\it A probabilistic graphical model (PGM) depicting the kind
of model discussed in this paper. Produced using daft ({\tt daft-pgm.org}).
The number, $N$, of components in the model is an unknown parameter, as are the
properties of the components. The data depends on the properties of the components.
Note that it is possible (and common) to have other parameters in the model
that point directly to the data. However, we have omitted such parameters
for simplicity.
\label{fig:pgm}}
\end{center}
\end{figure}

Here we have assumed the priors for $N$ and $\hyperparams$ are independent, and
the conditional prior for $\{\xx_i\}$ is iid and does not depend on $N$.
Given data $\mathcal{D}$, we usually want to calculate properties of the
posterior distribution, given by:
\begin{eqnarray}
p(N, \hyperparams, \{\xx_i\} | \mathcal{D}) \propto
p(N, \hyperparams, \{\xx_i\})
p(\mathcal{D} | N, \hyperparams, \{\xx_i\})\label{eq:bayes}
\end{eqnarray}
The structure of the model is depicted graphically in Figure~\ref{fig:pgm}.

The computational method used to calculate properties of the posterior distribution
is to generate samples from it using Markov Chain Monte Carlo (MCMC). Since
$N$ is unknown, an MCMC sampler needs to be able to jump between candidate
solutions with different numbers of components, using either birth-death MCMC
\citep{birthdeath} or the reversible jump framework of \citet{rjmcmc}.

The marginal likelihood, or evidence, is also an important quantity, and is
the normalisation constant of the posterior given in Equation~\ref{eq:bayes}:
\begin{eqnarray}
\mathcal{Z} &=& p(\mathcal{D})\\
&=& \sum_{i=0}^{N_{\rm max}} \int
p(N, \hyperparams, \{\xx_i\})p(\mathcal{D} | N, \hyperparams, \{\xx_i\})
\, d^N \xx_i \, d\hyperparams \label{eq:evidence}
\end{eqnarray}

Some authors have addressed these kinds of problems by
doing separate runs of Nested Sampling
(or another method of calculating marginal likelihoods),
with $N$ fixed at various trial values \citep[e.g.][]{fengji, feroz}.
Then, the posterior for $N$ can be calculated
based on the estimates of the marginal likelihood obtained as a function of $N$.
This approach does not generalise well to large $N$, as it would
require a large number of parallel runs.

In this paper, we introduce an approach that uses trans-dimensional MCMC moves
to infer $N$, but embeds this process in the Nested Sampling framework to
overcome difficulties that would be encountered by sampling the posterior
distribution directly. Our motivation
for using Nested Sampling is not primarily to calculate the marginal
likelihood $\mathcal{Z}$ (including summing over $N$, as in
Equation~\ref{eq:evidence}),
although $\mathcal{Z}$ is readily available from the output.
The motivation for using Nested Sampling, rather than just sampling the
posterior using trans-dimensional MCMC, is that the posterior often contains
difficult features, such as strong dependencies or multiple modes, that cause
problems with mixing.

\subsection{Phase Transitions and Nested Sampling}
In addition to the well-known challenges caused by multiple modes and strong
dependencies,
{\it phase transitions} \citep{skilling} can also cause difficulties for
Bayesian computation. Phase transitions occur when the posterior distribution
is a mixture of a ``slab'' with high volume but low density, and a ``spike''
with low volume but high density. To understand why this causes problems,
imagine sampling the posterior using the Metropolis algorithm. If the sampler
is in the slab, it will only rarely propose a move into the spike, because the
spike's volume is so small. If the sampler is in the spike, a proposal to move
into the slab will have a very low acceptance probability because the density
of the slab is so low relative to that of the spike. If the spike has a volume
$V_{\rm spike}$ and the slab has a volume $V_{\rm slab}$ then the number of
MCMC iterations needed to jump from the spike to the slab, or vice versa, is
of order $V_{\rm slab}/V_{\rm spike}$ which may be very large.

In many problems, the posterior distribution is a mixture of a spike component
and a slab component, however the total posterior probability in the slab
component is negligible. In this situation, posterior sampling is unaffected,
however problems can arise when calculating the marginal likelihood using
an annealing-type approach. For example, consider the thermodynamic integral,
which gives the log of the marginal likelihood:

\begin{eqnarray}
\log(\mathcal{Z}) &=& \int_0^1
\left<\log\left[L(\theta)\right]\right>_\beta \,d\beta
\end{eqnarray}

where $L(\theta)$ is the likelihood function, and the expectation is taken
with respect to the ``thermal'' distributions which are intermediate
between the prior $\pi(\theta)$ and the posterior:
\begin{eqnarray}
p_\beta(\theta) \propto \pi(\theta)L(\theta)^\beta.\label{eq:annealing}
\end{eqnarray}
If the posterior distribution is a mixture of a slab and a spike,
then for some range of inverse temperatures $\beta$ the probability mass in
the slab and the spike will be comparable. At these temperatures the MCMC
method will be unable to mix (the sampler will only visit the spike or the
slab, and not both) and will give an incorrect estimate of
$\left<\log\left[L(\theta)\right]\right>_\beta$.

Nested Sampling \citep{skilling} overcomes these problems by working with an
alternative sequence of probability distributions, proportional to the prior
but with a hard constraint on the value of the likelihood:
\begin{eqnarray}
p_{l}(\theta) \propto \pi(\theta)\mathds{1}\left[L(\theta) > l\right]\label{eq:constrained_prior}
\end{eqnarray}
The normalising constants of these distributions are given by
\begin{eqnarray}
X(l) &=& \int \pi(\theta)\mathds{1}\left[L(\theta) > l\right] \, d\theta\label{eq:X}
\end{eqnarray}
which is the amount of prior mass that has likelihood greater than
$l$. As Nested Sampling proceeds, $X$ shrinks geometrically with time, so
the number of iterations required to compress from the slab to the spike is
of order $\log(V_{\rm slab}/V_{\rm spike})$. In Diffusive Nested Sampling
(Section~\ref{sec:dns}), it takes of order $\log(V_{\rm slab}/V_{\rm spike})$
iterations to pass from the slab to the spike initially, and then
$\log(V_{\rm slab}/V_{\rm spike})^2$ time to revise this.
During this process Nested Sampling
visits states with likelihoods intermediate between those of the slab and the
spike, whereas no choice of $\beta$ in Equation~\ref{eq:annealing} will give
these states substantial probability.

\subsection{Diffusive Nested Sampling}\label{sec:dns}
The main difficulty with Nested Sampling is coming up with methods to sample
the constrained prior distributions of Equation~\ref{eq:constrained_prior}.
Various implementations of Nested Sampling exist, with different strategies
for sampling Equation~\ref{eq:constrained_prior}, as well as other features
to infer whether the run was successful, or to improve the accuracy of
the marginal likelihood estimate
\citep[e.g.][]{multinest, importance, diamonds, radfriends}.
Diffusive Nested Sampling
\citep[DNS][]{dnest} is an alternative version of Nested Sampling that is
appropriate when MCMC is used to explore the parameter space\footnote{
A C++ implementation of DNS, called {\tt DNest3}, is available at
{\tt https://github.com/eggplantbren/DNest3}}.

Rather than working with a sequence of constrained prior distributions like in
Equation~\ref{eq:constrained_prior}, DNS
replaces the posterior distribution with an alternative target
distribution composed of a {\it mixture} of the prior distribution with the
constrained priors, facilitating mixing between the multiple modes
and along degeneracy curves. The mixture of constrained priors is given by
a mixture of distributions like in Equation~\ref{eq:constrained_prior}:

\begin{eqnarray}
p(\theta) &=& \sum_{i=0}^n w_i
\frac{\pi(\theta)\mathds{1}\left[L(\theta) > l_i\right]}{X_i}
\end{eqnarray}
where $\{l_i\}$ is a set of likelihood thresholds or ``levels'',
constructed such that the associated prior mass $X_{i+1}$ of level
$i+1$ is approximately $e^{-1}$ times that of level $i$. The $\{w_i\}$ are a
set of mixture weights, usually chosen to emphasise the higher levels during
the initial stages of a run (while creating levels),
and set to uniform once the desired number of levels $n$ has been created.
Since DNS does not sample the posterior distribution, the samples need to be
assigned weights (in a manner similar to importance sampling) to represent
the posterior. See \citet{dnest} for more details about DNS.

\begin{figure}
\begin{minipage}[lr]{\textwidth}
\begin{center}
\includegraphics[width=2.7in]{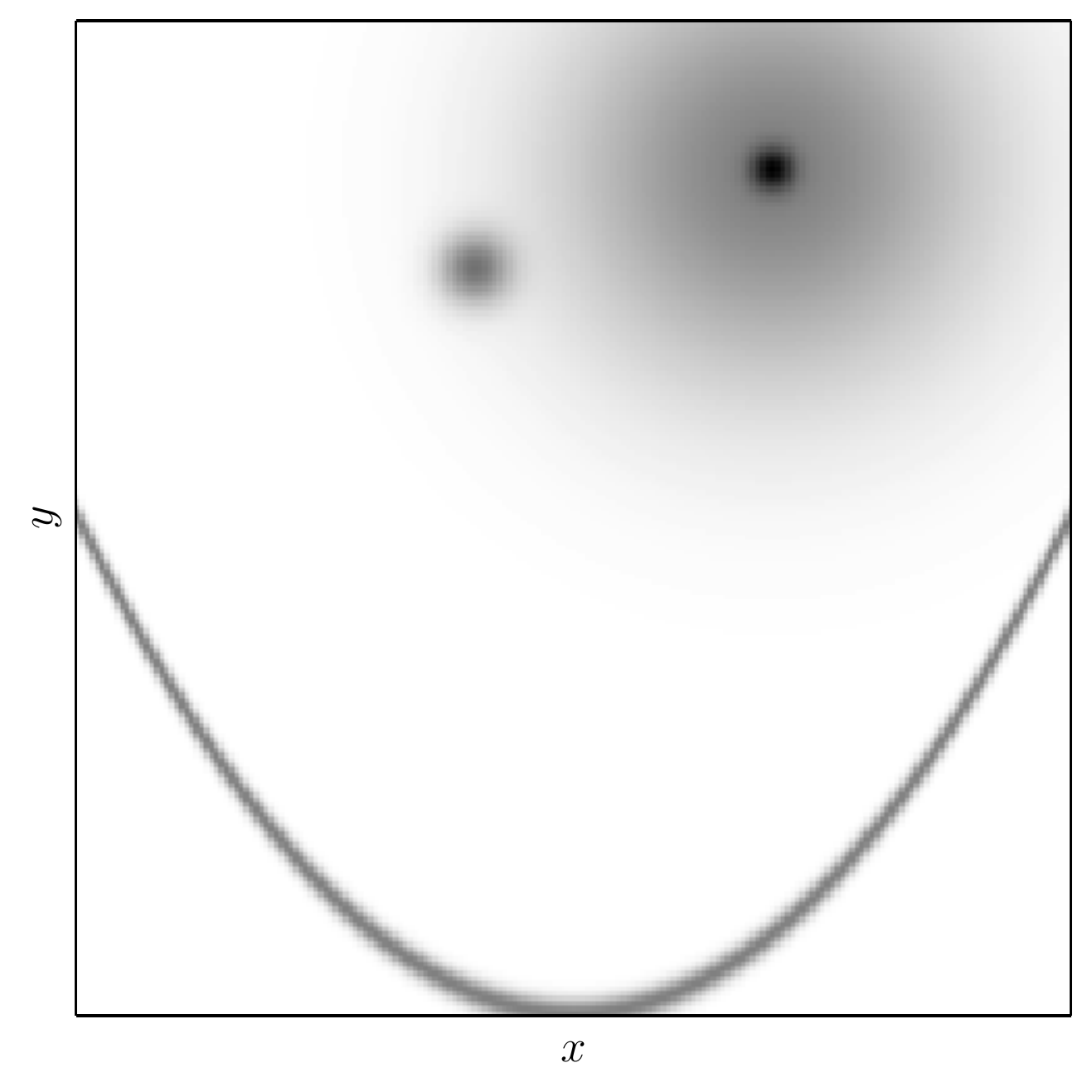}
\includegraphics[width=2.7in]{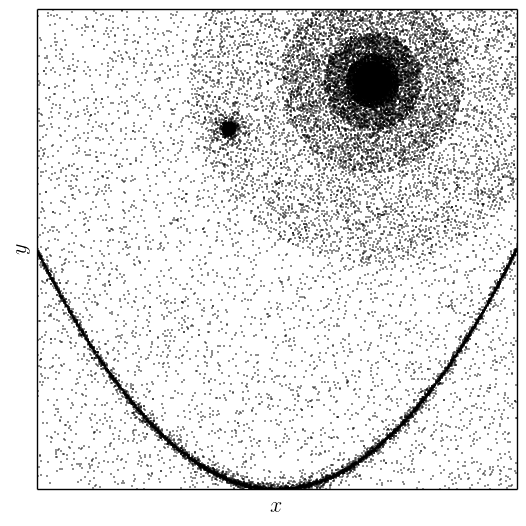}
\end{center}
\end{minipage}
\caption{\it {\bf Left}: An example of a two-dimensional posterior distribution
with multiple modes, strong dependence, and a phase transition (in the top
right mode). While this low dimensional example is not difficult to sample,
an analogous problem in high dimensions would be. {\bf Right}: Diffusive
Nested Sampling replaces the target posterior distribution by a mixture of
constrained priors. The density is non-negligible everywhere (since the prior
is a component of this distribution), and states intermediate between the two
phases in the top-right mode are appropriately up-weighted.
\label{fig:challenges}}
\end{figure}

The idea of replacing the target distribution with something other than the
posterior is similar to annealed importance sampling
\citep{neal} or parallel tempering \citep{pt},
where the target distribution is modified from the posterior to something
``easier''. However,
the sequence of distributions used in Nested Sampling avoids some of the
problems with annealing-based methods. In particular, there is no need to choose
a temperature schedule (a potentially large set of tuning parameters),
and the method does not fail when phase transitions
are present.
Figure~\ref{fig:challenges} shows a mock two dimensional posterior
density with three modes, one of which contains a strong dependence, and
another of which contains a phase transition. The target distribution used by
DNS includes the prior as part of the mixture, so travel between the modes is
possible.

Since DNS is effectively the Metropolis-Hastings algorithm applied to a
distribution other than the posterior, proposal distributions are needed
which define how a walker moves around the hypothesis space.
In Section~\ref{sec:proposals}, we outline a set of proposal distributions used
to explore the hypothesis space of possible values for $N$, $\hyperparams$,
and $\{\xx_i\}$.

A C++ implementation of the ideas described in this paper is available online
(under the terms of the GNU General Public Licence) at
{\tt https://github.com/eggplantbren/RJObject}.



\section{Metropolis proposals for the general problem}\label{sec:proposals}
We will now define a set of proposal distributions that can be used to
sample the prior distribution
$p(N, \hyperparams, \{\xx_i\}) = p(N) p(\hyperparams) \prod_{i=1}^N p(\xx_i | \hyperparams)$.
These proposals will need to satisfy detailed balance in order to be valid
when used inside DNS. When the proposal satisfies detailed balance with respect
to the prior, the DNS algorithm can incorporate hard likelihood constraints by
rejecting any proposal whose likelihood does not satisfy the constraint.
The overall proposal is a mixture of all of the proposals listed here.

When using DNS,
the proposals typically should be very heavy tailed. When
DNS is exploring a distribution close to the prior, large proposals will
generally be needed to explore the prior efficiently.
When DNS is exploring a distribution constrained to
very high values of the likelihood function, much smaller proposals will be
appropriate. Rather than attempting expensive tuning (since the number of
distributions involved may number in the hundreds or even thousands), we will
apply the heavy tailed proposal distributions and simply accept that there will
be some waste. If a parameter $u$ has a uniform prior between 0 and 1, then
a particular choice of heavy-tailed proposal is defined by:
\begin{eqnarray}
u' &:=& \textnormal{mod}\left(u + 10^{1.5 - 6a}b, 1\right)\label{eqn:heavytailed}
\end{eqnarray}
where $a \sim \textnormal{Uniform}(0, 1)$ and $b\sim \mathcal{N}(0,1)$.
This is like a standard gaussian/normal random walk proposal, but with a
non-fixed step size, resulting in a scale mixture of normals.
When
$a$ is close to zero, the size of the proposal is of order 10, which (because
of the mod function) effectively randomises $u'$ from the prior. If $a$ is
close to 1, the scale of the proposal is roughly $10^{-4}$ times the prior
width. The coefficient of $a$ was set to 6 because smaller proposals are
usually not necessary. The value 1.5 in the exponent was found to optimize the
performance of this proposal when sampling from $U(0, 1)$ and $\mathcal{N}(0,1)$
priors.

Most of the proposals discussed in this paper involve changing one
(or a subset) of the parameters and/or hyperparameters, while keeping the
others fixed. Whenever a heavy-tailed proposal distribution
is required, we use the proposal from Equation~\ref{eqn:heavytailed},
or an analogous one, e.g. a discrete version when the proposal is to change $N$.
The {\tt DNest3} package monitors the acceptance rate as a function of level.
In most applications, the acceptance probability is close to 1 (or may be
precisely 1) when sampling the prior, and decreases to 5-40\% when sampling
high levels.

\subsection{Proposals that modify $N$}\label{sec:proposal1}
The first kind of proposal we consider are proposals that change the
dimension of the model, i.e. proposals that change the value of $N$. By
necessity, these will also change the parameters $\{\xx_i\}_{i=1}^N$ because
the number of parameters will be changed. The proposals used here are
traditionally called {\it birth and death} proposals.

We will assume that the prior for $N$ is a discrete uniform distribution
between 0 and some $N_{\rm max}$, inclusive. The proposal starts by choosing
a new value $N'$ according to
\begin{eqnarray}
N' &=& \textnormal{mod}\left(N + \delta_N, N_{\rm max} + 1\right)
\end{eqnarray}
where $\delta_N$ is drawn from a heavy tailed
distribution which is a discrete analogue of Equation~\ref{eqn:heavytailed}
($N$ is treated as a real number in $[0, N_{\rm max} + 1]$ and then we take
the integer part at the end).
The most probable values of $\delta_N$ are $\pm 1$, but values
of order $N_{\rm max}$ also have some probability, to allow fast exploration
when the target distribution is similar to the prior. If $N' = N$, $N'$ is
regenerated as $N+1$ with probability 0.5, and $N-1$ with probability 0.5,
and then wrapped back into the range $\{0,...,N_{\rm max}\}$ if necessary.

If $N' > N$, i.e. the proposal is to add components to the model,
the extra parameters needed, $\{\xx_i\}_{i=N+1}^{N'}$,
are drawn from their prior conditional on the current value of the
hyperparameters, i.e. the conditional prior $p(\xx | \hyperparams)$.
If $N' < N$, i.e. the proposal is to remove components from the model,
then $N - N'$ components must be selected for removal. All
of the components have the same probability of being selected for removal.

\subsection{Proposals that modify components}\label{sec:proposal2}
We now consider a proposal distribution that modifies one or more of the
components $\{\xx_i\}$, while keeping the number of components $N$, as well as the
hyperparameters $\hyperparams$, fixed.

Let $F(x; \hyperparams)$ be a function that takes an component $x$ and transforms it
to a value $u$ that has a uniform distribution between 0 and 1, given $\hyperparams$.
If the component $x$ consists of a single scalar value, $F$ is the cumulative
distribution of the conditional prior. Denote the inverse of $F$ by $G$.

A proposal
for an component involves transforming its parameters to $[0, 1]$ using $F$,
making the proposal in terms of $u$, and then transforming back.
Specifically, the proposal chooses
a new value $\xx_i'$ from the current value $\xx_i$ as follows:
\begin{eqnarray}
u_i &:=& F(\xx_i; \hyperparams)\\
u_i' &:=& \textnormal{mod}\left(u_i + \delta_u, 1\right)\\
\xx_i' &:=& G(u_i'; \hyperparams).
\end{eqnarray}
where $\delta_u$ is drawn from a heavy-tailed distribution as in
Equation~\ref{eqn:heavytailed}.

Choosing just one component to change is most appropriate when the DNS
distribution is very constrained. When it is close to the prior, bolder
proposals that change more than one component at a time are possible. Hence,
we choose the number of components to change from a heavy tailed distribution
wherer the most probable value is 1 but there is also a nontrivial probability
of proposing to change $\sim N$ components at once.

\subsection{Proposals that change the hyperparameters,
keeping the components fixed}\label{sec:proposal3}
Another kind of proposal that we will consider is a proposal that keeps all of
the components fixed in place (i.e. leaves $\{\xx_i\}$) unchanged, but changes
the hyperparameter(s) from their current value $\hyperparams$
to a new value $\hyperparams'$. The proposal for the hyperparameter(s) is chosen
to satisfy detailed balance with respect to $p(\hyperparams)$ and should be
heavy-tailed.

The overall Metropolis acceptance probability
for this kind of move, if sampling the prior (or the constrained prior of DNS)
must include the following factor:
\begin{eqnarray}
\frac{\prod_{i=1}^N p(\xx_i | \hyperparams')}{\prod_{i=1}^N p(\xx_i | \hyperparams)}
\label{eqn:acceptance_prob}
\end{eqnarray}
Since this proposal leaves the components $\{\xx_i\}$ fixed, it will usually not
affect the value of the likelihood, and therefore the likelihood will not need
to be recomputed. Therefore this proposal is most useful for mixing the values
of the hyperparameters $\hyperparams$ when the target distribution is highly
constrained.

\subsection{Proposals that change the hyperparameters
and all of the components}\label{sec:proposal4}
The above proposals allow for changes to $N$, $\hyperparams$, and $\{\xx_i\}$,
and are therefore sufficient to allow for ``correct'' exploration of the
prior distribution, and the constrained prior distributions used in DNS.
However, they do not necessarily allow for {\it efficient} exploration, even
of the prior itself. The main reason for this is the inability for large
changes to be made to the hyperparameters $\hyperparams$. If the proposed change
from $\hyperparams$ to $\hyperparams'$ is large, the ratio in
Equation~\ref{eqn:acceptance_prob} is likely to be very small, and the move
will probably be rejected.

Therefore, we allow an additional move that changes $\hyperparams$ to a new
value $\hyperparams'$, but rather than leaving the components $\{\xx_i\}$ fixed,
they are ``dragged''
so they represent the distribution $p(\xx | \hyperparams')$ rather than
$p(\xx|\hyperparams)$. We do this by making use of the transformation functions
$F(x; \hyperparams)$ and $G(x; \hyperparams)$ defined in
Section~\ref{sec:proposal2}.

The ``dragging'' process works as follows, and must be carried out on
each component:
\begin{eqnarray}
u_i &:=& F(\xx_i; \hyperparams)\\
\xx_i' &:=& G(u_i; \hyperparams')
\end{eqnarray}
In other words, the components' parameters are transformed to $[0,1]$ using the
current value of the hyperparameters, and then transformed back using the
proposed value of the hyperparameters so they represent the new conditional prior
rather than the old one.

\section{Sinusoidal example}\label{sec:sinewaves}
In this section we demonstrate a seemingly simple example which exhibits
a phase transition, making a standard MCMC approach difficult. Suppose
a signal is composed of $N$ sinusoids, of different periods, amplitudes,
and phases. The signal is observed at various times $\{t_i\}_{i=1}^m$ with
noise, and we want to use the resulting data to infer the number of sinusoids
$N$, along with the periods $\{T_i\}_{i=1}^N$, amplitudes $\{A_i\}_{i=1}^N$,
and phases $\{\phi_i\}_{i=1}^N$.
This kind of model has many applications, and has been solved analytically
under certain sets of assumptions \citep[see e.g.][]{bretthorst, 2014arXiv1412.0467M}.
Similar
models have been used in many different fields
\citep[e.g.][]{2003AIPC..659....3B, 2005PhRvD..72b2001U, 2007ApJ...654..551B,
2009MNRAS.395.2226B}.
Note that this problem is also very closely related to the
problem of detecting exoplanet signals in radial velocity data, which has
attracted a lot of research attention in recent years
\citep[e.g.][]{gregory, fengji, 2011MNRAS.415.3462F}.

\subsection{Simulated Dataset}
\begin{figure}
\begin{center}
\includegraphics[scale=0.5]{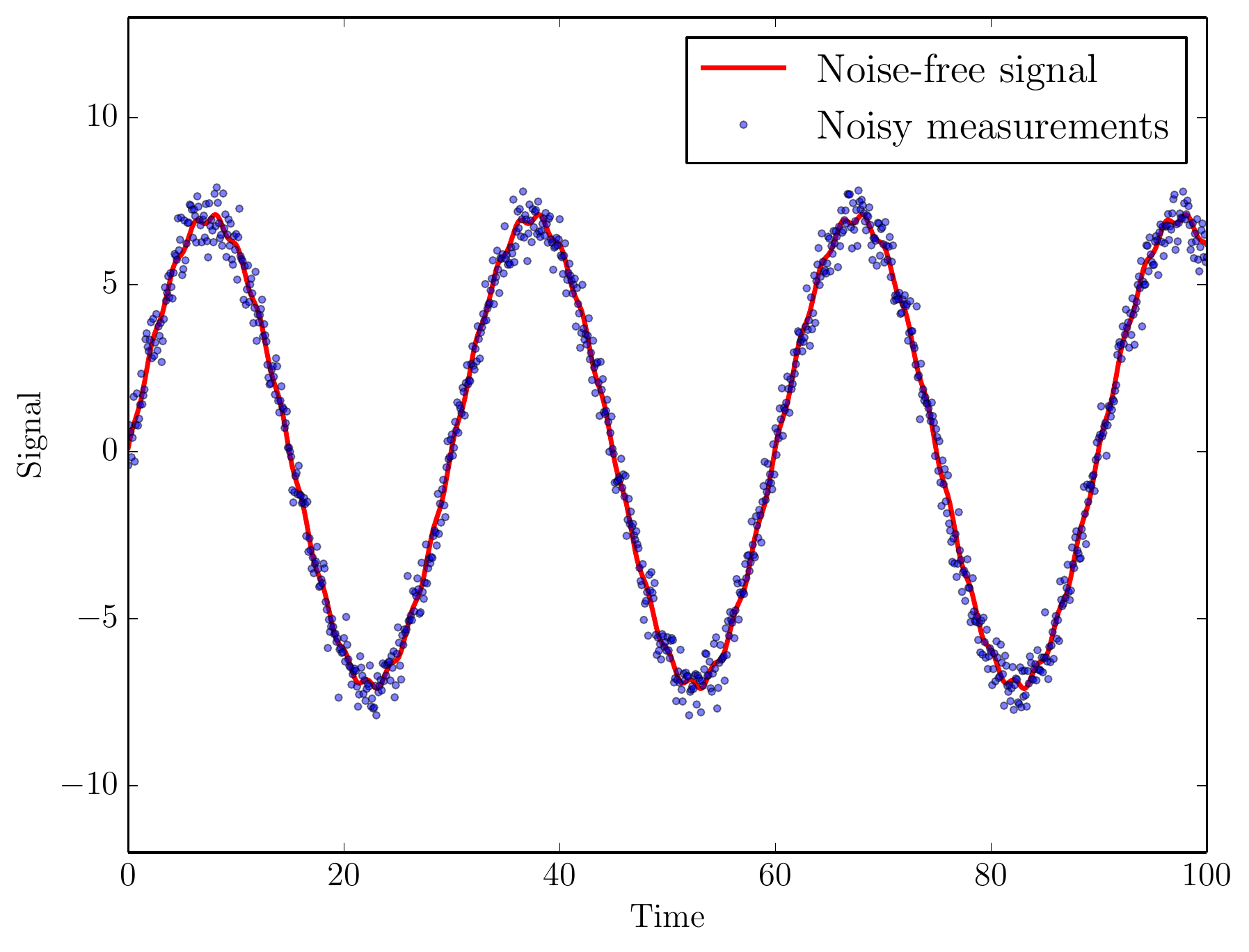}
\caption{\it The simulated data for the sinusoidal example. The solid line shows
the true signal which consists of a large slow oscillation and a much smaller
amplitude fast oscillation. The points are the measurements, simulated with a
noise standard deviation of $\sigma = 0.5$.
\label{fig:sinewave_data}}
\end{center}
\end{figure}

To demonstrate the techniques,
we simulated some data based on the assumption $N=2$, i.e. the true signal was
composed of two sinusoids.
The simulated data is shown in Figure~\ref{fig:sinewave_data}, and shows a
large, low period oscillation with a smaller, much faster oscillation
superimposed. The noise level is such that the fast oscillation is difficult
to detect. The true values of the parameters were
$N=2$, $\mathbf{A} = \{7, 0.155\}$,
$\mathbf{T}=\{30, 2\}$, and $\boldsymbol{\phi} = \{0, 1\}$. The signal was
observed at $n=1001$ points equally spaced between $t=0$ and $t=100$.
The true form of the signal is:
\begin{eqnarray}
y(t) = \sin\left(\frac{2\pi t}{30}\right) +
0.3 \sin\left(\frac{2\pi t}{2} + 1\right)
\end{eqnarray}
and the noise standard deviation was $\sigma = 0.5$.

\subsection{Prior Distributions}
To infer $N$, the number of sinusoids, as well as all the periods, amplitudes,
and phases, we need to define prior distributions.
In the notation of Section~\ref{sec:introduction}, the parameters of the
each component are
\begin{eqnarray}
\xx_i &=& \{A_i, T_i, \phi_i\}.
\end{eqnarray}
For the conditional prior $p(\xx_i | \hyperparams)$, we introduced a single
hyperparameter $\mu$, such that $A_i$ given $\mu$ has an exponential prior
with mean $\mu$. For the periods, we assigned a log-uniform distribution for
the periods between fixed boundaries, and a uniform distribution for the phases
between 0 and $2\pi$. In other words, our prior is only hierarchical for the
amplitudes $\{A_i\}$. The prior for $\mu$ was a log-uniform prior, with
probability density $\propto 1/\mu$, where $\mu \in [10^{-3}, 10^3]$.

The model for the shape of the (noise-free) signal is
\begin{eqnarray}
y(t) &=& \sum_{i=1}^N A_i \sin \left(\frac{2\pi t}{T_i} + \phi_i\right)
\end{eqnarray}
where there are $N$ sinusoids in the signal, the
amplitudes are $\{A_i\}$, the periods are $\{T_i\}$, and the phases are
$\{\phi_i\}$.
The sampling distribution (probability distribution for the data given the
parameters) was a normal (gaussian) distribution with mean zero and standard
deviation $\sigma$, applied independently to each data point:
\begin{eqnarray}
Y_i | N, \{A_i\}, \{T_i\}, \{\phi_i\}, \sigma \sim
\mathcal{N}\left(y(t_i), \sigma^2\right).
\end{eqnarray}
This also depends on an additional noise standard deviation parameter $\sigma$
which will also be inferred. We used a log-uniform prior for $\sigma$ between
$10^{-3}$ and $10^{3}$.

\subsection{Results}
\begin{figure}
\begin{center}
\includegraphics[scale=0.6]{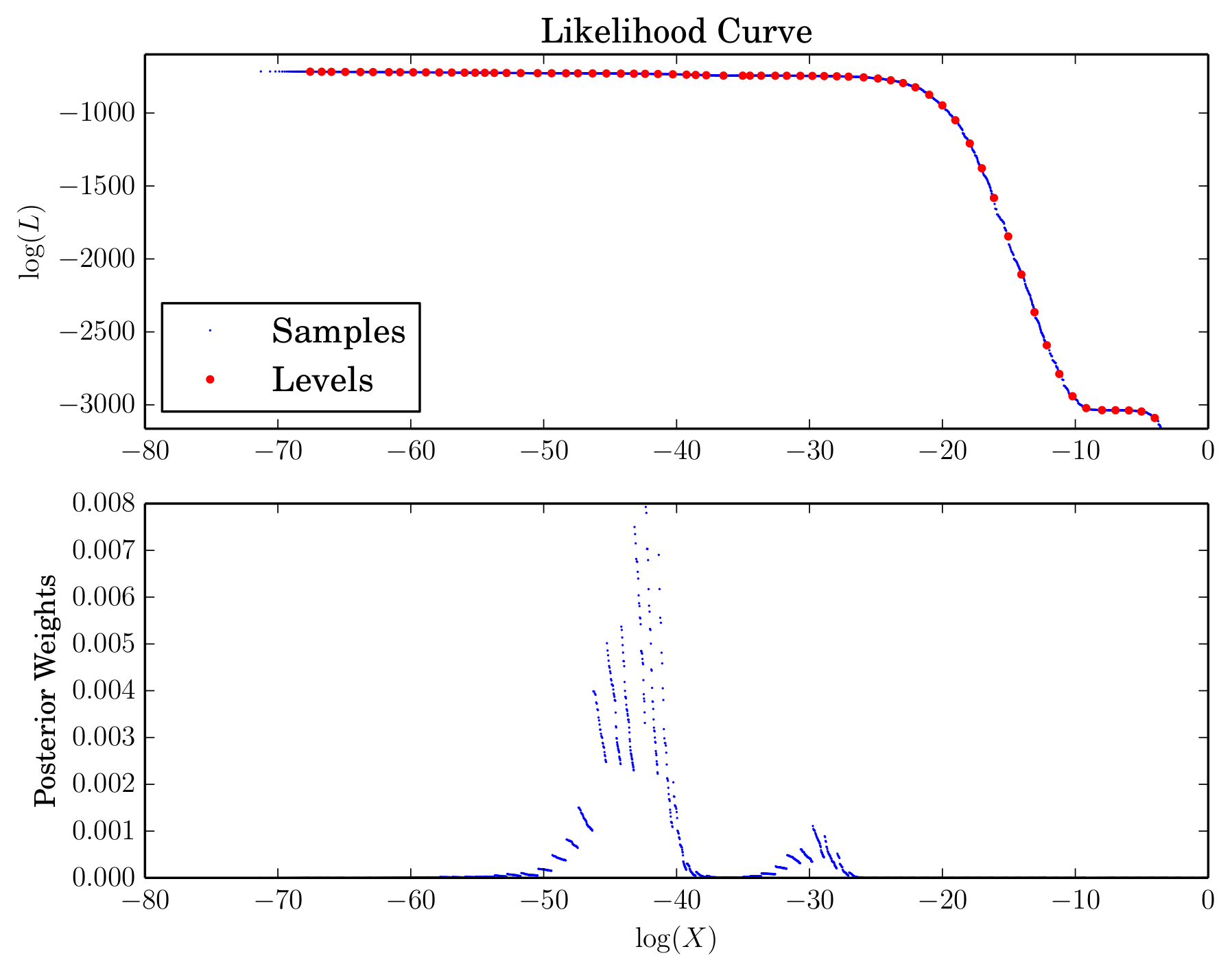}
\caption{\it {\bf Top panel: }
The shape of the likelihood function with respect to prior mass.
{\bf Bottom panel: }The posterior weights of the saved samples. The existence
of a phase transition causes this to display two separate peaks, which
would be difficult to mix between if we simply attempted to sample the posterior
distribution.
\label{fig:sinewaves_likelihood}}
\end{center}
\end{figure}

We ran DNS on the simulated data shown in Figure~\ref{fig:sinewave_data}.
The (log) likelihood as a function of (log) enclosed prior mass $X$
(Equation~\ref{eq:X}) is plotted in Figure~\ref{fig:sinewaves_likelihood}
This curve is a standard output of Nested Sampling methods and can be used to
gain insight into the structure of the problem. Particularly, when
phase transitions are present, this curve will have concave-up
regions \citep{skilling}.

Two phase transitions can be seen in Figure~\ref{fig:sinewaves_likelihood}. The
first phase transition, at $\log(X) \approx -10$ nats,
separates ``noise-only'' models (where the
entire dataset is accounted for by the noise term) from ``one-sinusoid'' models
where $N=1$. At $\log(X) \approx -35$ nats, a second phase transition separates
$N=1$ models from $N=2$ models. This phase transition causes a double-peaked
feature in the posterior weights (lower panel of
Figure~\ref{fig:sinewaves_likelihood}).

If we were to do this analysis by trying to explore the posterior directly,
rather than by using Nested Sampling, it would be difficult to jump between
the $N=1$ solution and the $N=2$ solution, because the posterior distribution
would be composed of a mixture of a small-volume but high-likelihood spike
and a large-volume but (relatively) low likelihood
slab. Note that this phase transition still exists if we condition on $N=2$,
so is not caused by the trans-dimensional model. To demonstrate this,
Figure~\ref{fig:trace_logl} shows the value of the log likelihood versus time
for 45 million iterations of MCMC targeting the posterior distribution.
The sampler should visit states with log likelihood $\sim -745$ about 10\%
of the time (corresponding with models that don't contain the small oscillation),
and the rest of the time should be spent in states with log likelihood around
$-730$. In this run transitions between these two phases occurred only a
handful of times. In a practical application, computing the posterior probability
for the existence of the small oscillation might be the whole point of the
calculation, so mixing between models where it is present and models where it
is absent is crucial.

\begin{figure}
\begin{center}
\includegraphics[scale=0.5]{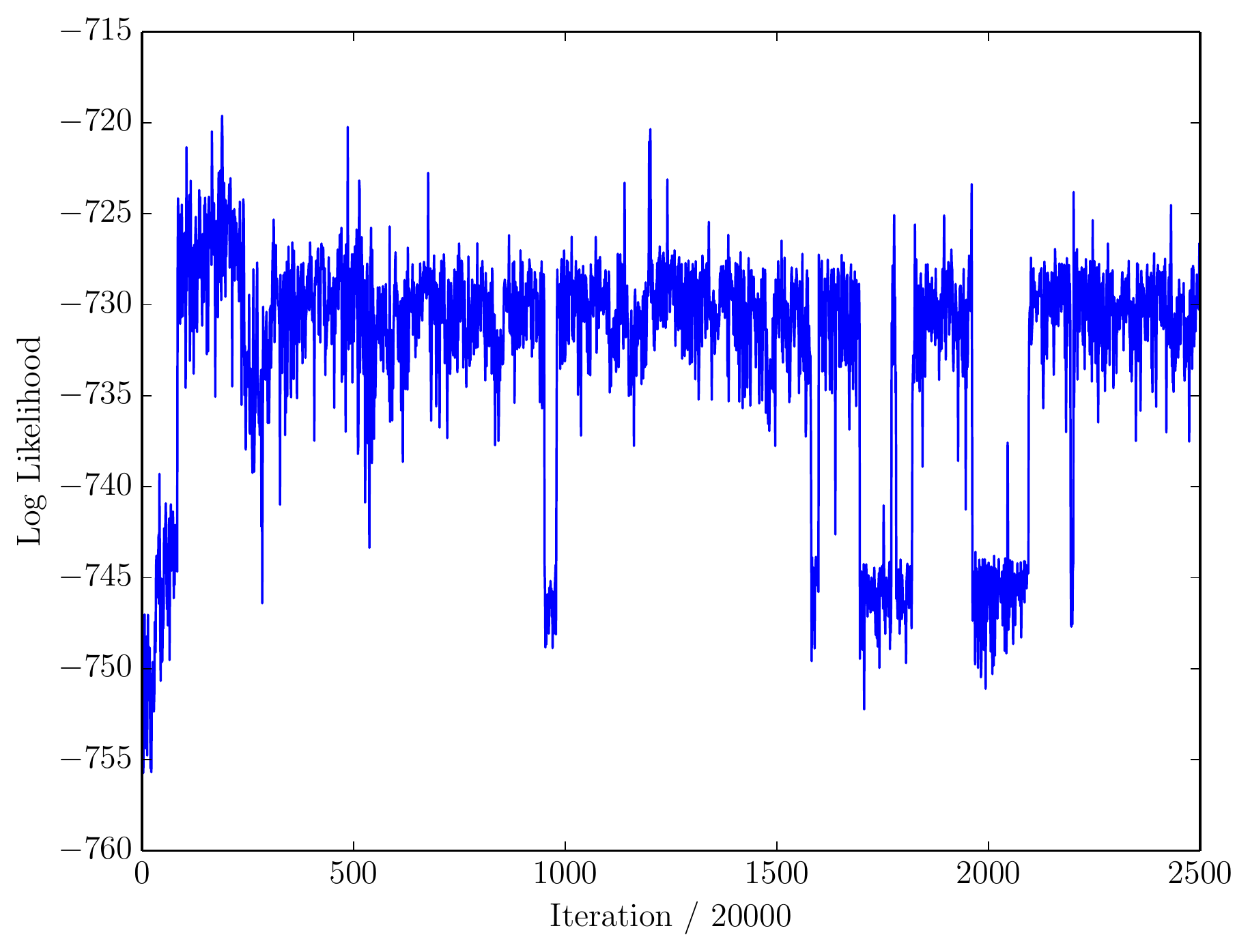}
\caption{\it A ``trace plot'' of the log likelihood in the sinusoidal model, for an
MCMC chain designed to sample the posterior distribution. The two phases
have log-likelihoods around -730 and -745 respectively, however transitions
between the two phases are quite uncommon.
\label{fig:trace_logl}}
\end{center}
\end{figure}

\begin{figure}
\begin{center}
\includegraphics[scale=0.5]{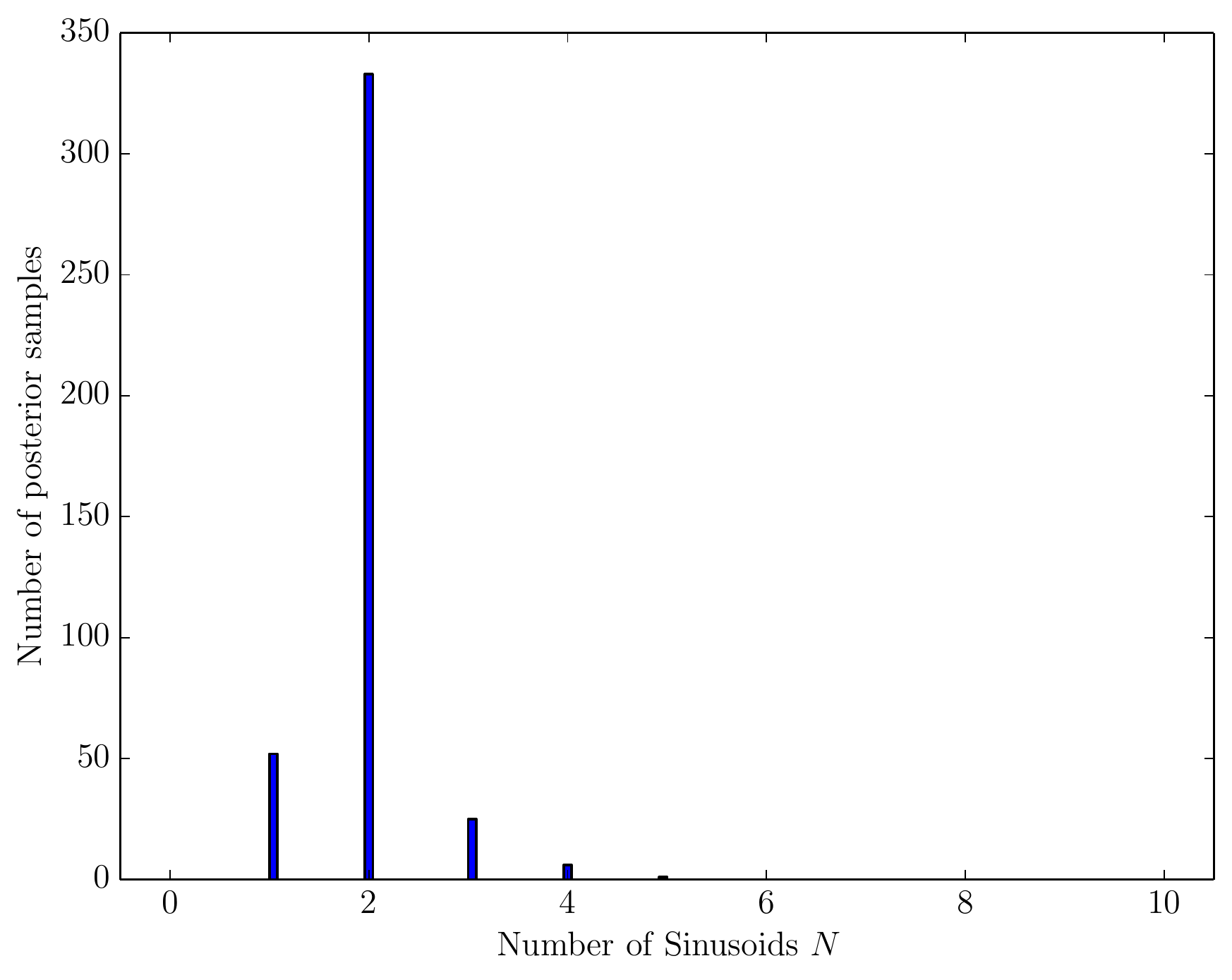}
\caption{\it The inference for $N$, the number of sinusoids, based on the
sinusoid data. The true value was $N=2$, which is also the most probable value
given the data. Of course, this is not always the case.
\label{fig:N_result}}
\end{center}
\end{figure}

Although DNS explores a distribution other than the posterior (a mixture of
constrained priors is used instead), the saved samples can be assigned
importance weights so we can represent the posterior distribution. The
marginal posterior distribution for $N$ is shown in Figure~\ref{fig:N_result}.
The most probable value is $N=2$ which is also the true value, and there is
a small probability for $N=1$. The decreasing probabilities for $N > 2$ are
due to the well-known natural ``Occam's Razor'' effect that can occur in
Bayesian Inference \citep[e.g.][Chapter 28]{mackay}.

The marginal likelihood estimate returned by DNS for this data was
$\log(\mathcal{Z}) = -771.8$, and the information, or Kullback-Leibler
divergence from the prior to the posterior, was estimated to be 39.8 nats.

\begin{figure}
\begin{center}
\includegraphics[scale=0.5]{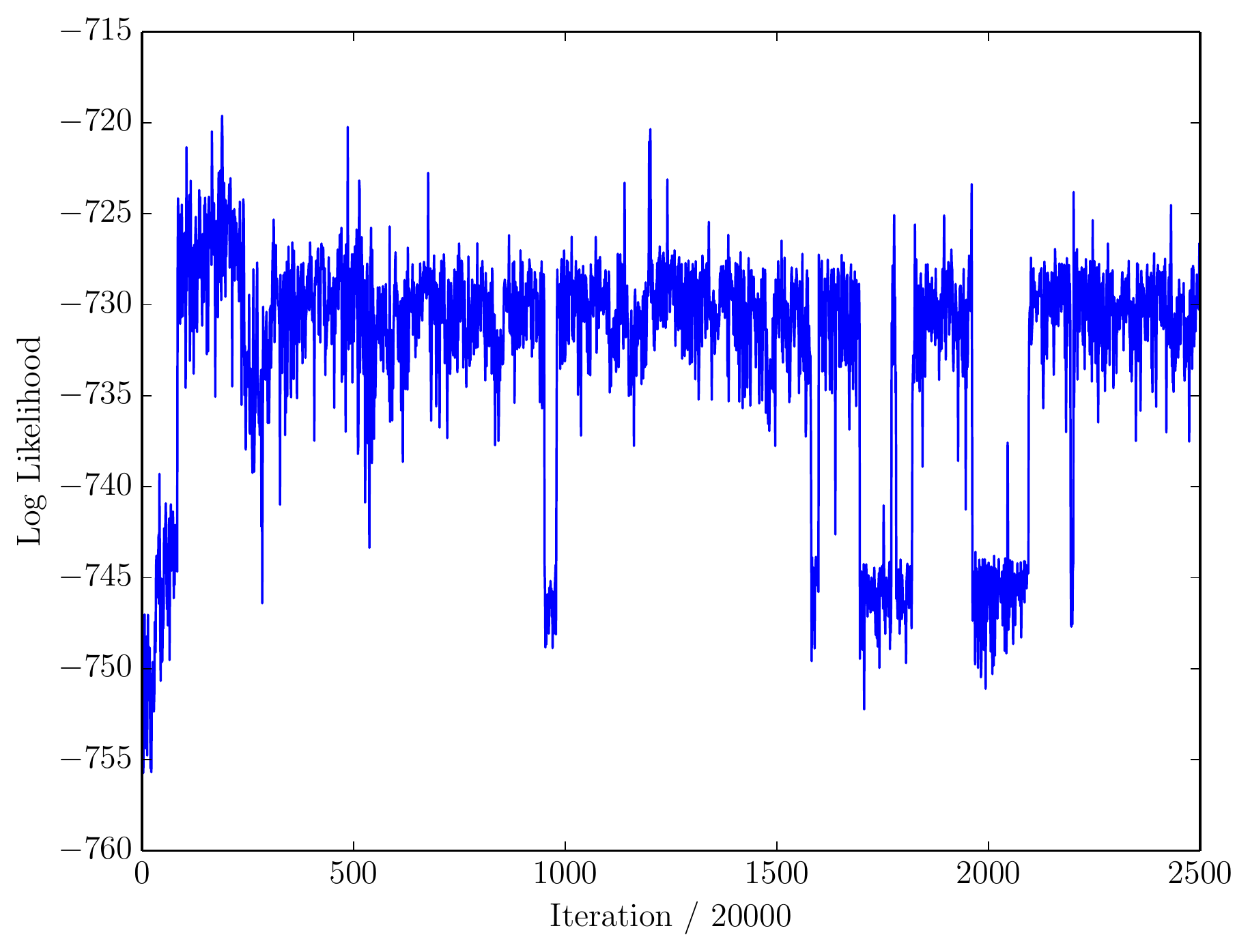}
\caption{\it A ``trace plot'' of the log likelihood in the SineWave model, for an
MCMC chain designed to sample the posterior distribution. The two phases
have log-likelihoods around -730 and -745 respectively, however transitions
between the two phases are quite uncommon.
\label{fig:trace_logl}}
\end{center}
\end{figure}

\section{``Galaxy Field'' Example}
Source detection is an important problem in astrophysics. Given some data,
usually one or more noisy, blurred, images of the sky, we would like to know
how many components $N$ (such as stars or galaxies) are in the image, and their
properties (such as flux, size, orientation). Various approaches exist,
covering a wide spectrum between ad-hoc approaches and principled inference
approaches
\citep[e.g.][]{irwin, sextractor, dolphot, 2003MNRAS.338..765H, starfield}.
The more principled techniques tend to be much more computationally
intensive, so should only be applied to interesting or
especially challenging subsets of astronomical imaging. In fact, the approach
used by \citet{starfield} was essentially an early version of the methodology
presented in this paper.

\begin{figure}
\begin{center}
\includegraphics[scale=0.7]{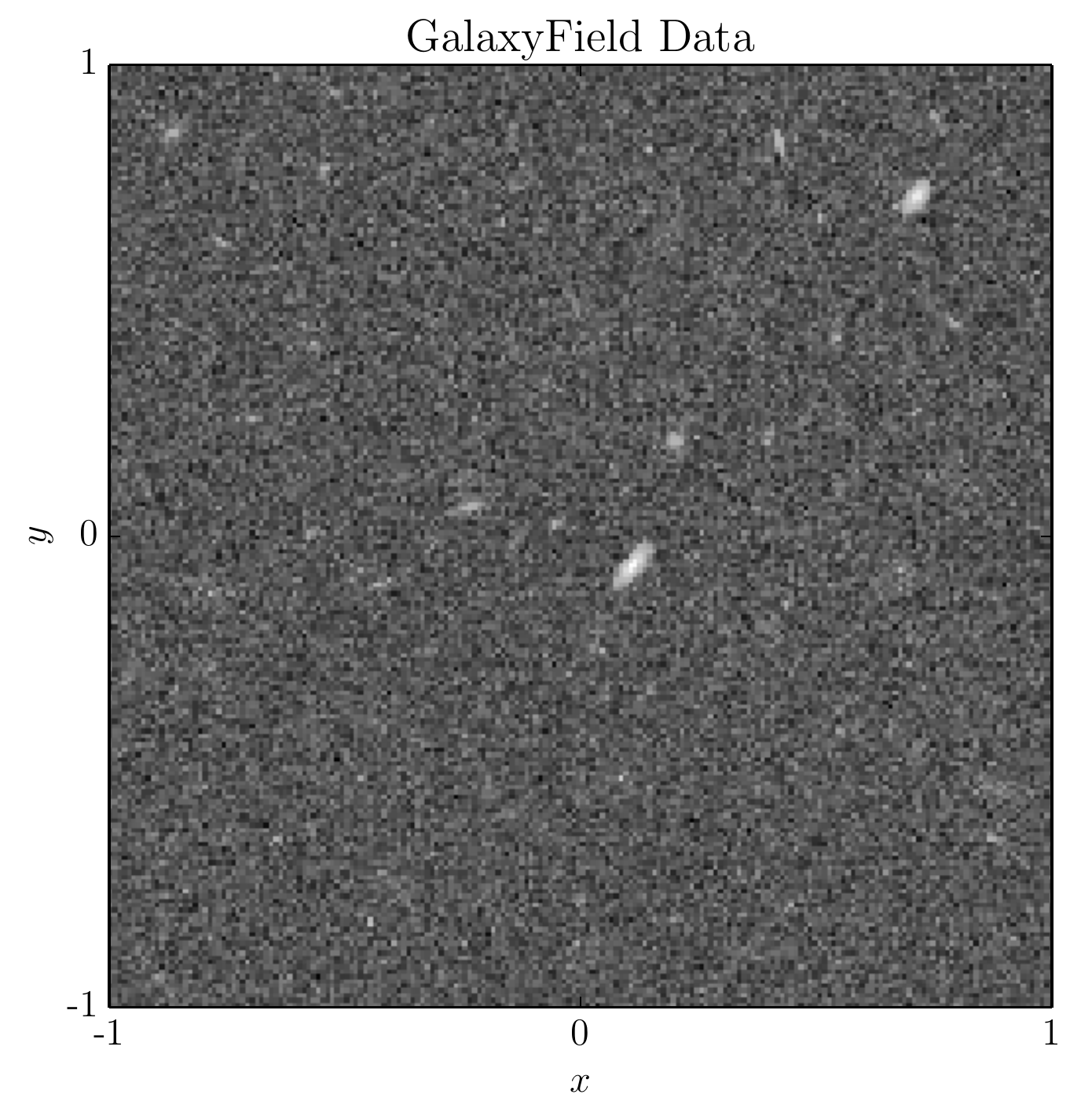}
\caption{\it The simulated image for the GalaxyField example. The image is
$200\times 200$ pixels in size and contains $N=47$ ``galaxies''.
\label{fig:galaxyfield_data}}
\end{center}
\end{figure}

In this section we apply
the DNS approach to a toy version of the problem of detecting and quantifying
extended components such as galaxies in a noisy image.
We model each galaxy as a mixture of two concentric elliptical gaussian
components, with
the following eight parameters. Firstly, two
parameters $x_c$ and $y_c$ describe the central position of the galaxy within
the image. The flux $f$ describes the total integral of the galaxy's
intensity profile. The axis ratio $q$ describes the ellipticity of the galaxy
and $\theta$ its orientation angle with respect to horizontal. The radius of the
bigger gaussian is a parameter $w$.
Finally
we include a ``radius ratio'' $u$ describing the ratio of the radius of the
smaller gaussian with respect to that of the bigger gaussian, and a parameter
$v$ describing the fraction of the total flux in the smaller gaussian
(therefore the fraction of the light in the bigger gaussian is $1-v$).
In a coordinate system aligned with the major axis of the ellipse, the
surface brightness profile of a ``galaxy'' is given by
\begin{eqnarray}
\rho(x, y) &=&
\frac{f(1-v)}{2\pi w^2}\exp\left[-\frac{1}{2w^2}\left(qx^2 + y^2/q\right)\right]\\
&&+\frac{fv}{2\pi (uw)^2}\exp\left[-\frac{1}{2(uw)^2}\left(qx^2 + y^2/q\right)\right].
\end{eqnarray}
We generated a simulated $200 \times 200$ pixel noisy image
(Figure~\ref{fig:galaxyfield_data}) containing 47 ``galaxies'',
in order to test the algorithm.

\subsection{Priors}
For the conditional prior for the total fluxes of the galaxies, and for the
widths of the galaxies, we used the Pareto
distribution. The probability density function for a variable $x$ given
hyperparameters $x_{\rm min}$ and $a$ is
\begin{eqnarray}
p(x | x_{\rm min}, a) &=&
\left\{
\begin{array}{lr}
\frac{a x_{\rm min}^{a}}{x^{a + 1}}, & x \geq x_{\rm min}\\
0, & \textnormal{otherwise.}
\end{array}
\right.
\end{eqnarray}
Since we are using this for both the fluxes and the widths of the galaxies,
there will be four hyperparameters in total: two lower limits (one for the
fluxes and one for the widths) and two slopes. The lower limits were assigned
log-uniform priors between $10^{-3}$ and $10^3$, and uniform priors between 0
and 1 were assigned to the {\it inverses} of the slopes. This latter choice
was based on the fact that a Pareto distribution for $x$ is an exponential
distribution for $\ln(x)$, with scale length $1/a$. If we do not expect the
scale length to be very extreme then we might want to restrict it to $[0,1]$
with a uniform prior.

The conditional priors for the radius ratios $u$ and the flux ratios $w$ were both
uniform distributions. Since $u \in [0,1]$ and $v \in [0,1]$, the upper limits
of the uniform conditional priors must be between 0 and 1. Therefore the prior
for the upper limit was a uniform distribution between 0 and 1, and the prior
for the lower limit given the upper limit was a uniform distribution between 0
and the upper limit.

As in Section~\ref{sec:sinewaves}
we allowed the noise standard deviation $\sigma$ to be a free parameter with
a log-uniform prior between $10^{-3}$ and $10^3$.

\subsection{Results}
\begin{figure}
\begin{center}
\includegraphics[scale=0.6]{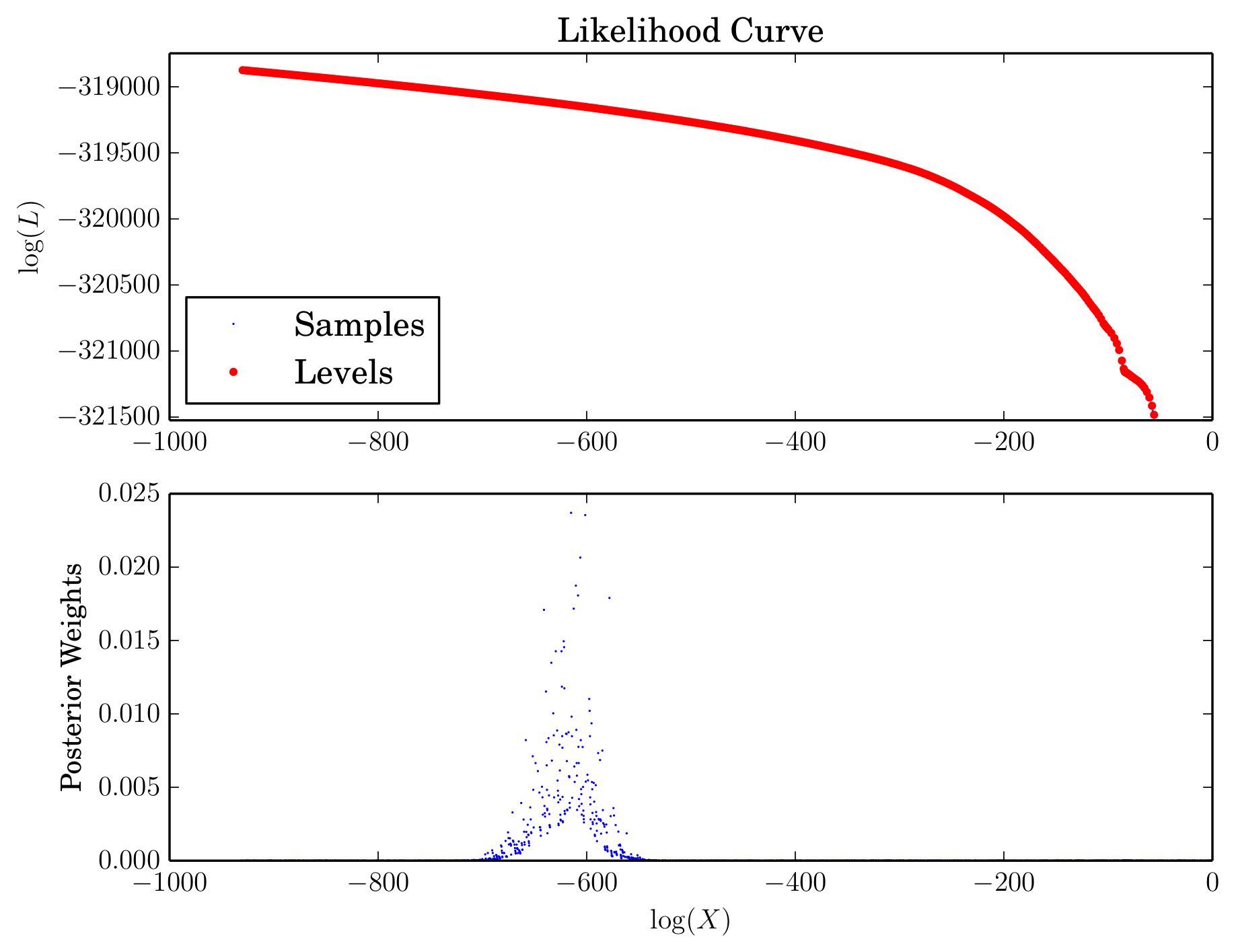}
\caption{\it {\bf Top panel: }
The shape of the likelihood function with respect to prior mass for the
GalaxyField data. Note the phase
transition at $\log(X) \approx -100$ nats. This separates models with just
the few brightest galaxies from models with many faint galaxies as well.
While this doesn't affect the
posterior distribution (unlike the sinewave example), it would affect calculation
of the marginal likelihood if annealing were used.
{\bf Bottom panel: }The posterior weights of the saved samples.
\label{fig:galaxyfield_likelihood}}
\end{center}
\end{figure}

\begin{figure}
\begin{center}
\includegraphics[scale=0.5]{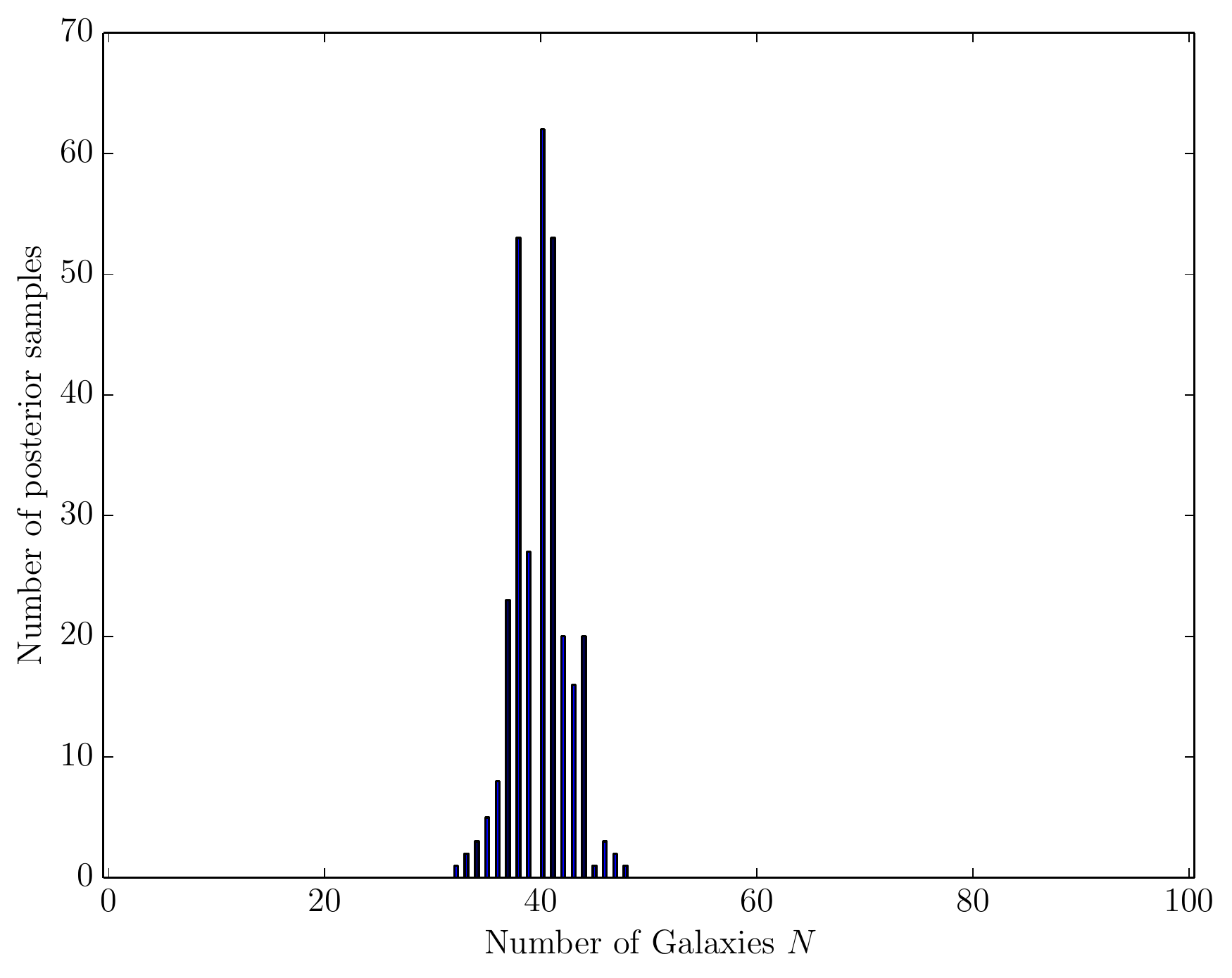}
\caption{\it The inference for $N$, the number of galaxies, based on the
GalaxyField data. The true value was $N=47$.
\label{fig:N_galaxy_result}}
\end{center}
\end{figure}

The results from running DNS on the simulated galaxy data are shown in
Figures~\ref{fig:galaxyfield_likelihood} and~\ref{fig:N_galaxy_result}.
This problem, like the sinusoidal problem, exhibits a phase transition, which
can be seen in Figure~\ref{fig:galaxyfield_likelihood}. The phase transition
separates models which only fit the brightest galaxies from models that
also fit the faint galaxies close to the noise level. This phase transition
does not affect the posterior distribution, as the low-$N$ solutions are
eventually found to have very low importance. However it would cause difficulty
for a thermal approach to calculating the marginal likelihood.

The marginal likelihood estimate returned by DNS for this data was
$\log(\mathcal{Z}) = -319707.6$, and the information, or Kullback-Leibler
divergence from the prior to the posterior, was estimated to be 550.2 nats.


\section{Optimisation Techniques}\label{sec:optimisation}
In both of the examples discussed in this paper, the likelihood evaluation
involved computing a mock noise-free signal from the current value of the
model parameters. The mathematical form of the mock signal was a sum over
all $N$ components present. Computing the mock signal is often the most
expensive step in the evaluation of the likelihood.

However, many of the proposal distributions used involve changing a subset
of the parameters, while keeping others fixed. Considerable speedups can be
achieved by, when appropriate, updating the mock signal to reflect the proposed
change to the model, rather than computing the entire mock signal from scratch.
For example, if the proposal is to add two new components to the model, the
mock signal can simply be updated by adding the effect of the two new components
to the model.

In general, it is possible to compute the number of components that have been
affected by the proposal. If this is greater than or equal to $N$, we compute
the mock signal from scratch, otherwise we subtract the effect of those
components that have been removed and add in the effect of those that have been
added. In the C++ implementation of this work, when a proposal takes
place, the difference between the old and new model is cached and is available
to the user, so that the likelihood can be updated rather than computed
from scratch. Using this technique offers a speed advantage of a factor of
$\sim 2$ for both of the example problems. The sinusoidal example could be
solved in a few minutes on a desktop computer, whereas the galaxy example took
much longer, about a day, to execute.

\section{Conclusions}
In this paper, we presented a general approach to trans-dimensional Bayesian
inference problems. To compute the posterior distribution in these contexts,
birth and death MCMC moves are commonly used. However, the approach presented
here replaces the posterior distribution with the alternative target
distribution used in Diffusive Nested Sampling. This is done to facilitate
mixing between multiple modes, along strong dependencies,
and between different phases. Additionally,
the marginal likelihood can be computed including the sum over the unknown
number of components. The method was demonstrated on two illustrative examples
inspired by astronomical data analysis, but should be applicable in other
contexts such as mixture modelling. Software (in C++) implementing the
general Metropolis proposals for these models, as well as the specific
examples, is available at {\tt https://github.com/eggplantbren/RJObject}.

\section*{Acknowledgements}
This work is supported by a Marsden Fast-Start grant
from the Royal Society of New Zealand. I would like to thank the following
people for valuable conversations and inspiration:
Anna Pancoast (UCSB), David Hogg (NYU), Daniel Foreman-Mackey (NYU),
Courtney Donovan (Auckland), Tom Loredo (Cornell), Iain Murray (Edinburgh),
Ewan Cameron (Oxford),
John Skilling (MaxEnt Data Consultants), and Daniela Huppenkothen
(Amsterdam, NYU).


\begin{thebibliography}{}
\bibitem[Bertin and
Arnouts(1996)]{sextractor} Bertin, E., Arnouts, S.\ 1996.\ SExtractor: Software for source extraction.\ Astronomy and Astrophysics Supplement Series 117, 393-404.

\bibitem[Bretthorst(1988)]{bretthorst} Bretthorst, G.~Larry.\  1988.\
Bayesian Spectrum Analysis and Parameter Estimation.\ In Lecture Notes in
Statistics, 48, Springer-Verlag, New York, New York.


\bibitem[Bretthorst(2003)]{2003AIPC..659....3B} Bretthorst, G.~L.\ 2003.\ 
Frequency Estimation, Multiple Stationary Nonsinusoidal Resonances With 
Trend.\ Bayesian Inference and Maximum Entropy Methods in Science and 
Engineering 659, 3-22. 


\bibitem[Brewer et al.(2007)]{2007ApJ...654..551B} Brewer, B.~J., Bedding, 
T.~R., Kjeldsen, H., Stello, D.\ 2007.\ Bayesian Inference from 
Observations of Solar-like Oscillations.\ The Astrophysical Journal 654, 
551-557. 

\bibitem[Brewer and Stello(2009)]{2009MNRAS.395.2226B} Brewer, B.~J., 
Stello, D.\ 2009.\ Gaussian process modelling of asteroseismic data.\ 
Monthly Notices of the Royal Astronomical Society 395, 2226-2233. 

\bibitem[Brewer et al.(2013)]{starfield} Brewer, B.~J., 
Foreman-Mackey, D., Hogg, D.~W.\ 2013.\ Probabilistic Catalogs for Crowded 
Stellar Fields.\ The Astronomical Journal 146, 7. 

\bibitem[\protect\citeauthoryear{Brewer, P{\'a}rtay,
\& Cs{\'a}nyi}{2011b}]{dnest} Brewer B.~J., P{\'a}rtay L.~B., Cs{\'a}nyi G., 2011,
Statistics and Computing, 21, 4, 649-656. arXiv:0912.2380

\bibitem[Brewer et al.(2011)]{2011MNRAS.412.2521B} Brewer, B.~J., Lewis,
G.~F., Belokurov, V., Irwin, M.~J., Bridges, T.~J., Evans, N.~W.\ 2011.\
Modelling of the complex CASSOWARY/SLUGS gravitational lenses.\ Monthly
Notices of the Royal Astronomical Society 412, 2521-2529

\bibitem[Buchner (2014)]{radfriends}
Buchner, Johannes.~A statistical test for Nested Sampling algorithms.~Statistics
and Computing (2014): 1-10.

\bibitem[Corsaro and De 
Ridder(2014)]{diamonds} Corsaro, E., De Ridder, J.\ 2014.\ DIAMONDS: A new Bayesian nested sampling tool. Application to peak bagging of solar-like oscillations.\ Astronomy and Astrophysics 571, AA71.

\bibitem[Dolphin(2000)]{dolphot} Dolphin, A.~E.\ 2000, PASP, 
112, 1383 

\bibitem[Feroz et al.(2013)]{importance} Feroz, F., Hobson, M.~P., 
Cameron, E., Pettitt, A.~N.\ 2013.\ Importance Nested Sampling and the 
MultiNest Algorithm.\ ArXiv e-prints arXiv:1306.2144. 

\bibitem[\protect\citeauthoryear{Feroz, Hobson,
\& Bridges}{2009}]{multinest} Feroz F., Hobson M.~P., Bridges M., 2009, MNRAS, 398, 1601

\bibitem[Feroz et al.(2011)]{2011MNRAS.415.3462F} Feroz, F., Balan, S.~T.,
Hobson, M.~P.\ 2011.\ Detecting extrasolar planets from stellar radial
velocities using Bayesian evidence.\ Monthly Notices of the Royal
Astronomical Society 415, 3462-3472.

\bibitem[Feroz and Hobson(2014)]{feroz} Feroz, F., Hobson, 
M.~P.\ 2014.\ Bayesian analysis of radial velocity data of GJ667C with 
correlated noise: evidence for only two planets.\ Monthly Notices of the 
Royal Astronomical Society 437, 3540-3549. 

\bibitem[\protect\citeauthoryear{Green}{1995}]{rjmcmc}
Green, P.~J., 1995, Reversible Jump Markov Chain Monte Carlo Computation and Bayesian Model Determination, Biometrika 82 (4): 711–732.

\bibitem[Gregory(2011)]{gregory} Gregory, P.~C.\ 2011.\ 
Bayesian exoplanet tests of a new method for MCMC sampling in highly 
correlated model parameter spaces.\ Monthly Notices of the Royal 
Astronomical Society 410, 94-110.

\bibitem[Hansmann(1997)]{pt} Hansmann, Ulrich HE., 1997, ``Parallel tempering algorithm for conformational studies of biological molecules.'', Chemical Physics Letters 281, no. 1 (1997): 140-150.

\bibitem[Hobson 
\& McLachlan(2003)]{2003MNRAS.338..765H} Hobson, M.~P., \& McLachlan, C.\ 2003, MNRAS, 338, 765 

\bibitem[Hou et al.(2014)]{fengji} Hou, F., Goodman, J., Hogg, 
D.~W.\ 2014.\ The Probabilities of Orbital-Companion Models for Stellar 
Radial Velocity Data.\ ArXiv e-prints arXiv:1401.6128.

\bibitem[Irwin(1985)]{irwin} Irwin, M.~J.\ 1985, MNRAS, 214,
575

\bibitem[Jasra et al(2005)]{label_switching} Jasra, A., Holmes, C.~C, and
Stephens, D.~A., Markov Chain Monte Carlo Methods and the Label Switching
Problem in Bayesian Mixture Modeling, Statistical Science
Vol. 20, No. 1, pp. 50-67

\bibitem[MacKay(2003)]{mackay} MacKay, David J.~C., 2003, ``Information theory, inference, and learning algorithms''. Vol. 7.
Cambridge: Cambridge university press, 2003.

\bibitem[Mortier et al.(2014)]{2014arXiv1412.0467M} Mortier, A., Faria, 
J.~P., Correia, C.~M., Santerne, A., Santos, N.~C.\ 2014.\ BGLS: A Bayesian 
formalism for the generalised Lomb-Scargle periodogram.\ ArXiv e-prints 
arXiv:1412.0467

\bibitem[Neal(2001)]{neal} Neal, R.~M., 2001, 
Annealed importance sampling, Statistics and Computing, vol. 11, pp. 125-139.

\bibitem[\protect\citeauthoryear{Skilling}{1998}]{massinf}
Skilling J., 1998, Massive Inference and Maximum Entropy, in Maximum Entropy
and Bayesian Methods, Kluwer Academic Publishers, Dordrecht/Boston/London p.14

\bibitem[\protect\citeauthoryear{Skilling}{2006}]{skilling} Skilling, J., 2006, Nested Sampling for General Bayesian Computation, Bayesian Analysis 4, pp. 833-860.

\bibitem[Stephens(2000)]{birthdeath} Stephens, M., 2000, ``Bayesian analysis of mixture models with an unknown number of components-an alternative to reversible jump methods.'', Annals of Statistics (2000), 40-74.

\bibitem[Umst{\"a}tter et al.(2005)]{2005PhRvD..72b2001U} Umst{\"a}tter, 
R., Christensen, N., Hendry, M., Meyer, R., Simha, V., Veitch, J., 
Vigeland, S., Woan, G.\ 2005.\ Bayesian modeling of source confusion in 
LISA data.\ Physical Review D 72, 022001. 

\end{thebibliography}
\end{document}